\documentclass[pteplogo]{ptephy_v2}

\preprintnumber{XXXX-XXXX} 
\usepackage{hyperref}

\usepackage{url} 
\usepackage{mathtools}

\begin{document}

\title{Binding energy of \texorpdfstring{$^{3}_{\Lambda}\rm{H}$}] and \texorpdfstring{$^{4}_{\Lambda}\rm{H}$}] via image analyses of nuclear emulsions using deep-learning}


\author[1,2*]{A. Kasagi}
\author[2,3,4*]{T. R. Saito}
\author[2,5]{V. Drozd}
\author[2]{H. Ekawa}
\author[2,6]{S. Escrig}
\author[2,7,8]{Y. Gao}
\author[2,9]{\\Y. He}
\author[2,7,8]{E. Liu}
\author[2]{A. Muneem}
\author[2,10]{M. Nakagawa}
\author[2,11,12]{K. Nakazawa}
\author[6]{\\C. Rappold}
\author[2]{N. Saito}
\author[1]{M. Taki}
\author[2]{Y. K. Tanaka}
\author[2]{H. Wang}
\author[2,4]{A. Yanai}
\author[13]{\\J. Yoshida}
\author[14]{M. Yoshimoto}

\affil[1]{Graduate School of Artificial Intelligence and Science, Rikkyo University, 3-34-1 Nishi Ikebukuro, Toshima-ku, Tokyo 171-8501, Japan}
\affil[2]{High Energy Nuclear Physics Laboratory, Cluster for Pioneering Research, RIKEN, 2-1 Hirosawa, Wako, Saitama 351-0198, Japan}
\affil[3]{GSI Helmholtz Centre for Heavy Ion Research, Planckstrasse 1, D-64291 Darmstadt, Germany}
\affil[4]{Department of Physics, Saitama University, Saitama, 338-8570, Japan}
\affil[5]{Energy and Sustainability Research Institute Groningen, University of Groningen, Groningen, Netherlands}
\affil[6]{Instituto de Estructura de la Materia, CSIC, Madrid, Spain}
\affil[7]{Institute of Modern Physics, Chinese Academy of Sciences, 509 Nanchang Road, Lanzhou, 730000, Gansu Province, China}
\affil[8]{University of Chinese Academy of Sciences, Beijing 100049, China}
\affil[9]{School of Nuclear Science and Technology, Lanzhou University, 222 South Tianshui Road, Lanzhou, Gansu Province, 730000, China}
\affil[10]{Department of Mathematics, Physics, and Computer Science, Faculty of Science, Japan Women’s University, Mejirodai 2-8-1, Bunkyo-ku, Tokyo 112-8681, Japan}
\affil[11]{Faculty of Education, Gifu University, 1-1 Yanagido, Gifu 501-1193, Japan}
\affil[12]{The Research Institute of Nuclear Engineering (RINE), University of Fukui, 1-3-33 Kanawa, Tsuruga, Fukui 914-0055, JAPAN}
\affil[13]{International Center for Synchrotron Radiation Innovation Smart, Tohoku University, Sendai 980-8572, Japan}
\affil[14]{RIKEN Nishina Center, RIKEN, 2-1 Hirosawa, Wako, Saitama 351-0198, Japan\email{ayumi.kasagi@rikkyo.ac.jp}\email{takehiko.saito@riken.jp}}

\begin{abstract}%
Subatomic systems are pivotal for understanding fundamental baryonic interactions, as they provide direct access to quark-level degrees of freedom. 
In particular, the inclusion of a strange quark introduces strangeness as a new dimension, offering a powerful tool for exploring nuclear forces. 
The hypertriton, the lightest three-body hypernuclear system, provides an ideal testing ground for investigating baryonic interactions and quark behavior involving up, down, and strange quarks.
However, experimental measurements of its lifetime and binding energy, key indicators of baryonic interactions, exhibit significant deviations in results obtained from energetic collisions of heavy-ion beams. 
Identifying alternative pa
thways for precisely measuring the hypertriton's binding energy and lifetime is thus crucial for advancing experimental and theoretical nuclear physics. 
Herein, we present an experimental study on the binding energies of $^{3}_{\Lambda}\rm{H}$(hypertriton) and $^{4}_{\Lambda}\rm{H}$, performed through the analysis of photographic nuclear emulsions using state-of-the-art technologies. 
By incorporating deep-learning techniques, we uncovered systematic uncertainties in conventional nuclear emulsion analysis and established a refined calibration protocol for accurately determining binding energies.
Our results are independent of those obtained from heavy-ion collision experiments, thereby offering a complementary measurement and opening new avenues for investigating few-body hypernuclei interactions.
\end{abstract}

\subjectindex{D14, H41, H42}

\maketitle

\section{Introduction}
Subatomic nuclei, including hyperons, baryons containing strange quarks, offer exceptional opportunities to investigate the fundamental mechanisms underlying nuclear forces and astrophysical phenomena.
Strangeness serves as a potent tool for probing short-range repulsive nucleon interactions at distances of ~1~fm or less, revealing processes that are difficult to observe in systems composed solely of up and down quarks \cite{nuclearforce_1, nuclearforce_2}. 
Detailed experimental data and theoretical studies of hyperon–nucleon interactions under flavor-SU(3) symmetry can elucidate the nature of three-body forces involving strangeness \cite{strangeness_3body} and provide insights into the internal structure of neutron stars, particularly in the context of recent gravitational wave discoveries \cite{LIGO}. 
At extreme densities, hyperons may appear in neutron star cores, influencing heavy-element synthesis via the rapid neutron-capture process and contributing to the explanation of how some neutron stars exceed two solar masses \cite{2ms_1, 2ms_2, 2ms_3, 2ms_4, 2ms_review, Strangeness_in_NS_1, Strangeness_in_NS_2}. 
Accordingly, hypernuclear research on systems composed of nucleons and hyperons serves as a pivotal link between nuclear physics and astrophysics.

Over the past 70 years, experimental studies have evolved from manual scanning of nuclear emulsions to missing-mass spectroscopy using real-time detectors and particle beams \cite{Hashimoto}.
Gamma-ray spectroscopy of excited states, decay-pion spectroscopy, and hyperon scattering have been employed to probe $\Lambda N$-interactions \cite{Yamamoto, MAMI_2016, Miwa}.
Nevertheless, for many types of single-$\Lambda$-hypernuclei, the primary data still stem from emulsion-based measurements of the 1970s.

A prominent example is the “hypertriton”, the lightest known hypernucleus, comprising a $\Lambda$ hyperon bound to a deuteron ($pn$). 
As a three-body $\Lambda NN$ system, it requires comprehensive understanding of hyperon–nucleon interactions and serves as a benchmark in hypernuclear studies.
In the 1960s, measurements using nuclear emulsion techniques reported a $\Lambda$-binding energy of $0.13 \pm 0.05\rm{(Stat.)}$ MeV ~ \cite{Juric}. 
In contrast to the proton–neutron pair in a deuteron, which has a total binding energy of $\sim2$~MeV, the $\Lambda$–deuteron system has a significantly lower binding energy, indicating a more loosely bound state. 
Consequently, the hypertriton lifetime, 263~ps for a free $\Lambda$ in vacuum, was commonly assumed to remain largely unchanged because of its minimal interaction with the deuteron \cite{PDG, lifetime_bindingenergy_1}.

However, interest in hypertriton properties was renewed in the 2010s when heavy-ion experiments reported lifetimes significantly shorter than that of a free $\Lambda$ \cite{HypHI, STAR_2010, ALICE_2016, STAR_2018, ALICE_2019}. 
Theoretical models struggled to reconcile these observations with the low binding energy of $0.13$~MeV, leading to the “hypertriton lifetime puzzle”. 
Subsequent measurements revised the average hypertriton lifetime to $237^{+9}_{-10}$ps \cite{STAR_2022, ALICE_2023, HypernuclearDataBase}. 
In 2020, the STAR collaboration reported a binding energy of $0.41 \pm 0.12\rm{(Stat.)} \pm 0.11\rm{(Syst.)}$~MeV, prompting re-evaluation of both lifetime and earlier binding energy estimates \cite{STAR_2020}. 
More recently, the ALICE collaboration reported a binding energy of $0.102 \pm 0.063\rm{(Stat.)}\pm 0.067\rm{(Syst.)}$~MeV, consistent with previous emulsion-based studies \cite{ALICE_2023}. 
Theoretical efforts have emphasized the correlation between lifetime and binding energy, further underscoring the need for more precise data \cite{lifetime_bindingenergy_1, lifetime_bindingenergy_2, lifetime_bindingenergy_3, lifetime_bindingenergy_4}.

Here, we report the detection of hypertriton events and high-precision binding energy measurements, achieved by combining nuclear emulsion analysis with deep-learning based image-style translation and object detection techniques.
The feasibility of this approach was demonstrated by detecting $^{3}_{\Lambda}\rm{H}$ and $^{4}_{\Lambda}\rm{H}$ events without requiring real training data \cite{Saito2021}.
The submicrometer spatial resolution of nuclear emulsion enables event-by-event identification, effectively eliminating background contamination and offering precision levels that surpass those of other experimental techniques.

A major challenge in binding energy measurements lies in ensuring the accuracy of the range–energy relation, which converts particle track-lengths into kinetic energies. 
Conventionally, this relationship is calibrated using monoenergetic $\alpha$-particles. 
However, in our study, a significant inconsistency emerged when comparing the calibration derived from $\alpha$-particles with that obtained from monoenergetic $\mu^{+}$ tracks. 
Specifically, the momenta of back-to-back emitted particles from the at-rest two-body decay of ${}^{4}_{\Lambda}\mathrm{H}$ deviated by $3.9\sigma$ from the expected values. 

To address this issue, we employed ATIMA, a modern stopping-power calculation tool that accounts for elemental composition effects, enabling accurate calibration of the range–energy relation for the E07 nuclear emulsion.
Using this revised calibration approach, we determined the binding energies as $B_{\Lambda} = 0.23 \pm 0.11(\mathrm{Stat.}) \pm 0.05(\mathrm{Syst.})$~MeV for $^{3}_{\Lambda}\mathrm{H}$ and $2.25 \pm 0.10(\mathrm{Stat.}) \pm 0.06(\mathrm{Syst.})$~MeV for $^{4}_{\Lambda}\mathrm{H}$. 
These results represent the first quantitative estimates of systematic uncertainties in modern hypernuclear binding energy measurements using nuclear emulsions. They mark a significant advancement in high-precision studies of various hypernuclei, contributing to a deeper understanding of baryonic interactions involving strangeness.

\section{Methods}
This study on the binding energy of $^{3}_{\Lambda}\rm{H}$ was conducted by analyzing decay events in nuclear emulsion sheets from the J-PARC E07 experiment \cite{E07_proposal}.
These emulsions, along with a diamond target placed in front of them, were exposed to secondarily charged particles, including $K^{-}$, at the K1.8 beamline \cite{K18_beamline}.
Originally, the experimental analysis concept relied on real-time detectors positioned around the nuclear emulsion to identify $p(K^{-}, K^{+})\Xi^{-}$ reactions occurring in the target. 
This approach considerably narrows the search area in the emulsion and  reduces the workload, enabling efficient detection of double-strangeness hypernuclei \cite{Nakazawa_book}. 
Concurrently, numerous single-$\Lambda$ hypernuclei, including hypertriton, are presumed to form within the emulsion via the elementary process $n(K^{-}, \pi^{-})\Lambda$, wherein the $\Lambda$ particles bind to simultaneously-emitted nuclear fragments. 
However, real-time detectors provide no information regarding the particles produced in association with the creation and decay of single-$\Lambda$ hypernuclei.
This limitation has become a key challenge in the search for $^{3}_{\Lambda}\rm{H}$ events.

\begin{figure*}[ht]
\centering
\includegraphics[width=140mm]{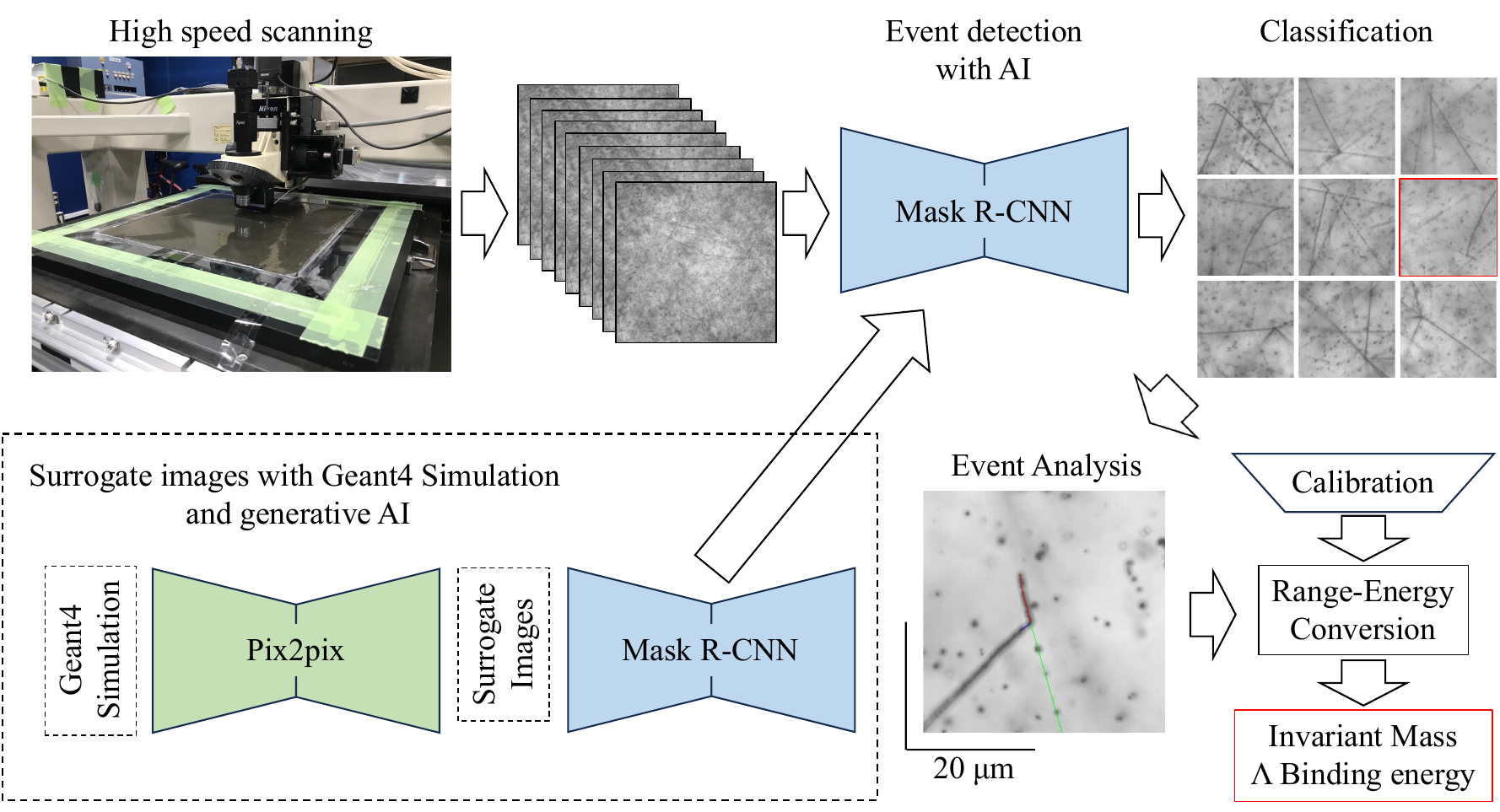}
\caption{{\bf Flow chart for binding energy analysis using nuclear emulsion and deep-learning}
A dedicated microscope (shown in the upper left) converts film-recorded tracks into digital images. 
To address the scarcity of real training data for rate events, we used Geant4 simulations and a deep-learning based style-transfer model to generate surrogate images.
These surrogate images enable the creation of an object-detection model that generalizes to actual data. 
The trained model is then applied to real microscopic images to identify candidate hypernuclear decay events. 
After classification—via manual inspection or other methods—the lengths and angles of associated tracks are measured. 
For each identified hypernuclear decay event, the calibrated Range–Energy relation is applied to convert track length into kinetic energy, allowing the invariant mass and $\Lambda$ binding energy to be determined.}
\label{fig:flow_chart}
\end{figure*}

To resolve this, we leveraged deep-learning’s generalization capabilities in object detection and image generation \cite{Saito2021}.
Recent advances in these domains have enhanced AI’s performance beyond human capabilities in certain classification and detection tasks, enabling the development of generative models that produce highly realistic images from textual or pictorial inputs.
By integrating these approaches, we established a unified framework to generate training data for the dedicated models.
Figure \ref{fig:flow_chart} presents an overview and the corresponding flowchart of the entire process. 
The following subsections describe each step of the development in detail.

\subsection{High-speed scanning system}
The critical components of the microscope were upgraded to enable fast image scanning, facilitating digitization of all E07 emulsion sheets within a few years.
The microscope stage for E07 is based on a design by UNIOPT Co., Ltd., and provides horizontal motion across a $380 \times 380~\mathrm{mm}^2$ area (stepping motor: Oriental Motor PK566-NBC) and vertical movement along the $z$-axis (stepping motor: Oriental Motor PK543-NBC).
Three key improvements accelerated the scanning process:
1. A 4-megapixel complementary metal-oxide-semiconductor (CMOS) camera operating at 150~fps replaced the previous charge-coupled device (CCD) camera, enabling rapid imaging over a large field of view.
2. A 20$\times$ Nikon air lens ($N.A.=0.35$) with a pixel size of $\sim 0.29\,\mathrm{\mu m}$ was introduced, which was equivalent to that of a 50$\times$ oil lens on a CCD camera.
3. These enhancements preserved both a wide field of view and sufficiently high resolution.

Faster imaging and a larger field of view was exploited to further optimize the stage motion. 
Images at multiple focal planes were acquired by moving only along the $z$-axis, capturing ~80 images of a $250\,\mathrm{\mu m}$-thick layer in $3\,\mathrm{\mu m}$ pitch. 
These sequential images enabled the reconstruction of three-dimensional track positions and angles. Although the camera can operate at 150~fps, the stepping motors cannot maintain stable high-speed motion. 
Hence, a piezo actuator was employed along the $z$-axis to ensure reliable and rapid focal adjustments. 
With the stage motor enabling $xy$ travel, the piezo actuator controlling focus, and integrated software management, the potential of the high-speed camera was fully realized \cite{HTS}.
The details of the microscope systems are provided in Table~\ref{tab:new_stage}.

\begin{table*}[htbp]
    \centering
    \caption{Comparison of the components and performances of microscopic stages for scanning\label{tab:new_stage}}
    \begin{tabular}{c|c|c}
    &Previous &Current\\
    \hline
    \hline
    Camera&CCD&CMOS\\
    \hline
    Objective lens&50$\times$, $\rm{NA} = 0.85$, (Oil)&20$\times$, $\rm{NA} = 0.35$, (Air)\\
    \hline
    View size [$\rm{pixel}^{2}$]&$512 \times 440$&$2048 \times 2048$\\
    \hline
    [$\rm{\mu m}^{2}$]&$138.2 \times 118.8$&$593.9 \times 593.9$\\
    \hline
    Frame rate [Hz]&50&150\\
    \hline
    Device for driving z-axis &Pulse Motor&Piezoelectric actuator\\
    \hline
    Speed [$\rm{m^{2}/h}$]&$4.5 \times 10^{-5}$&$8.3 \times 10^{-4}$ \\
    \hline
    [TB/day]&0.65~TB&12~TB
    \end{tabular}
\end{table*}

\subsection{Surrogate images with Geant4 Simulation and Generative AI}
A workflow was established to generate training datasets for rare hypernuclear decay events, for which the available real data were insufficient to develop a deep-learning model.
Using Geant4, a well-studied simulator for particle–nuclear physics, we modeled the sensitive volume of the nuclear emulsion and recorded charged-particle trajectories in 0.2\,$\mu\mathrm{m}$ steps \cite{Geant4}. 

Next, each particle’s coordinates were transformed into “hits” corresponding to the $\sim 1\,\mu\mathrm{m}$-diameter silver grains observed in the emulsion, matching the microscope’s specifications. 
Out-of-focus objects appear blurred; hence, these signals were projected onto an RGB color space: red indicates defocus from lower Z-planes, green corresponds to in-focus grains, and blue represents defocus from higher Z-planes. 
In addition, background tracks arising from unrelated particles or spallation reactions induced by the ${K}^{-}$ beam were superimposed using threshold-processed real images and Jet AA Microscopic transport model (JAM)-based model calculations, embedding the ${}^{3}_{\Lambda}\mathrm{H}$ signal in a more realistic visual context \cite{JAM}.

\begin{figure}[ht]
\centering
\includegraphics[width=70mm]{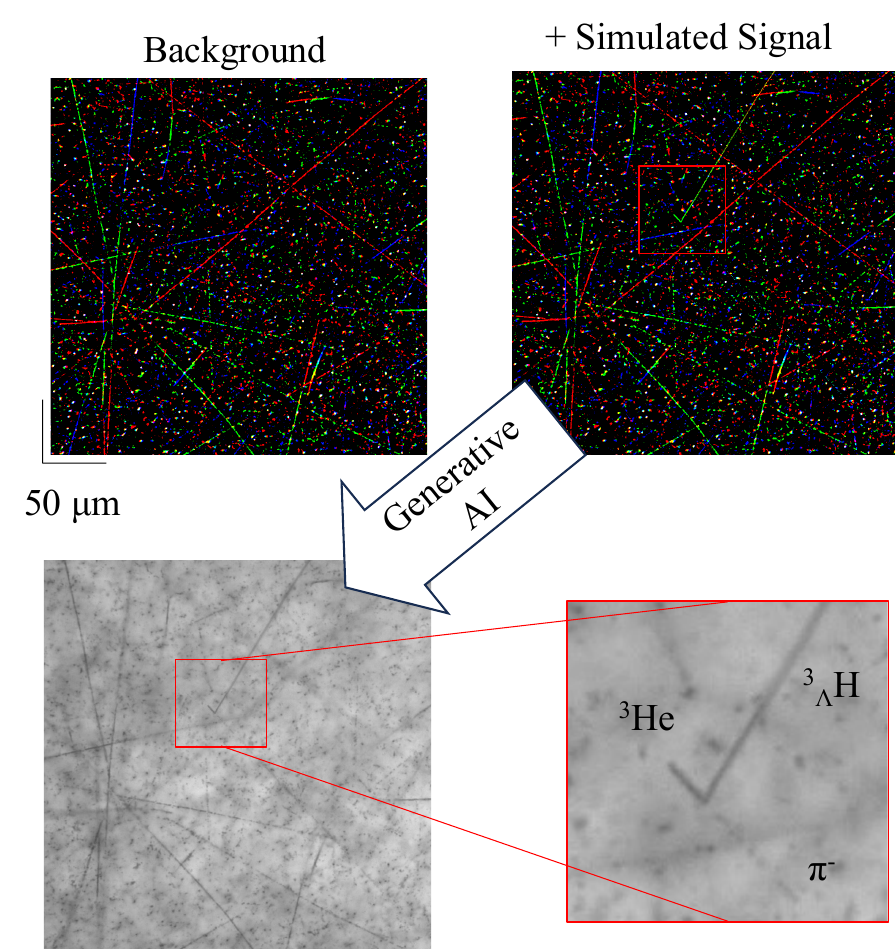}
\caption{{\bf Strategy for training data generation using image processing and generative AI.}\\
A hypernuclear decay event generated by Geant4 was overlaid onto a threshold-processed image from real data to simulate background events. 
The pix2pix model, trained separately for style transformation, renders the final image in a realistic “microscope-like” fashion. 
The magnified view highlights the incoming hypernucleus and the two daughter particle tracks, as observed in the nuclear emulsion.}
\label{fig:pix2pix}
\end{figure}

Nevertheless, these simulated images still differed markedly from actual microscope data. 
To bridge this gap and ensure that our object detection model generalized effectively to both simulated and real images, we employed pix2pix, an image-to-image translation network based on generative adversarial networks (GANs) \cite{GAN, pix2pix}. 
We trained pix2pix using paired data from the original and processed microscope images.
Finally, ${}^{3}_{\Lambda}\mathrm{H}$ decay events generated by Geant4 were converted into a “microscope-like” style, thereby creating surrogate training images for rare-event detection in nuclear emulsion. (See Figure \ref{fig:pix2pix})

\subsection{Event detection with AI}
\begin{figure}[ht]
\centering
\includegraphics[width=140mm]{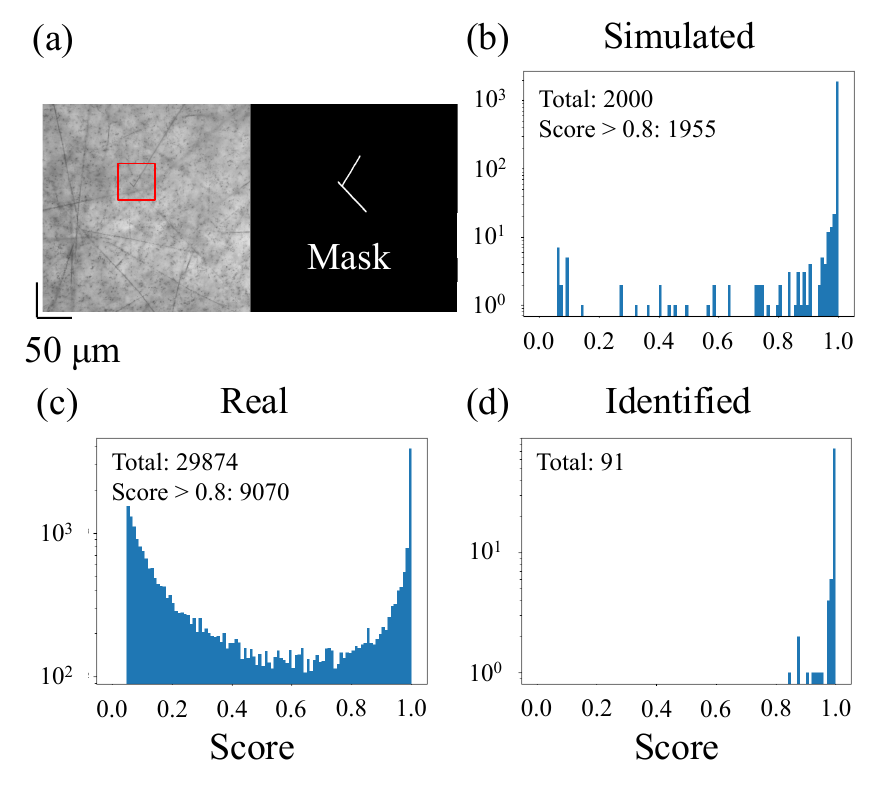}
\caption{{\bf Inference results from the dedicated model for detecting hypernuclear decay events.}
Panel (a) shows a sample input microscopic image (left) and the model’s output mask (right), indicating the segmentation and detection of a candidate for a two-body decay event. 
Panels (b), (c), and (d) display the distribution of detection scores for 2,000 simulated test images, ~30,000 events in two million real microscope images, and the hypernuclear events identified through our search and analysis, respectively. 
In both the simulated and real images, hypernuclear events were detected with scores close to the maximum value of 1.}
\label{fig:score_distribution}
\end{figure}

We trained a deep-learning based object detection model to identify two-body hypernuclear decay events, leveraging the simulated images described above. 
For this architecture, we adopted Mask R-CNN, which was state-of-the-art at the start of this project, and performed transfer learning (supervised fine-tuning) using ImageNet-pretrained weights \cite{MaskRCNN, ImageNet}. 
As the simulation pipeline provides precise positional and mask information for each hypernuclear decay, the manual annotation, typically required for object-detection training, could be bypassed. 
In total, ~10,000 images were used for training.
The Mask R-CNN outputs a classification likelihood (score) between 0 and 1 for each detected object. 
Figure \ref{fig:score_distribution}(a) presents a representative inference result, while panels (b), (c), and (d) show the score distributions for three distinct datasets: simulated test images independent of the training data, two million real microscope images, and two-body hypernuclei decay events identified by our analysis.
Validation of unused images demonstrated that many events yielded scores close to one. 

Conventionally, the detection threshold is determined from the receiver operating characteristic (ROC)-curve by examining true positive and false positive rates across varying threshold values. 
However, as estimating the number of positive samples for hypernuclear events, which are rare in real data, is difficult, the score threshold for candidate events was set to 0.8. This value was determined based on the results of tests on $\alpha$-decay, for which abundant real data are available \cite{Kasagi2023}.
Inference was then performed on approximately two million images from a $5 \times 5 \times 0.05~\rm{cm}^{3}$ emulsion volume, yielding ~10,000 candidate events with scores above the threshold. 
Of these, approximately 50 were selected as potential hypernuclear two-body decay events for detailed microscopic examination. 
Manual inspection confirmed the presence of hypernuclear events among these candidates; the scores for the identified hypernuclear events were similar to one, consistent with those observed in the simulated validation data.
Once a sufficient number of two-body hypernuclear decay events had been identified, this manual classification process was partially replaced by an independent CNN-based image classification model, further reducing the visual workload \cite{CNN_yoshida}.
On average, $1.92 \pm 0.06$ two-body hypernuclear decay events were identified per $5 \times 5 \times 0.05~\rm{cm}^{3}$ volume of emulsion.

\begin{figure}[ht]
\centering
\includegraphics[width=70mm]{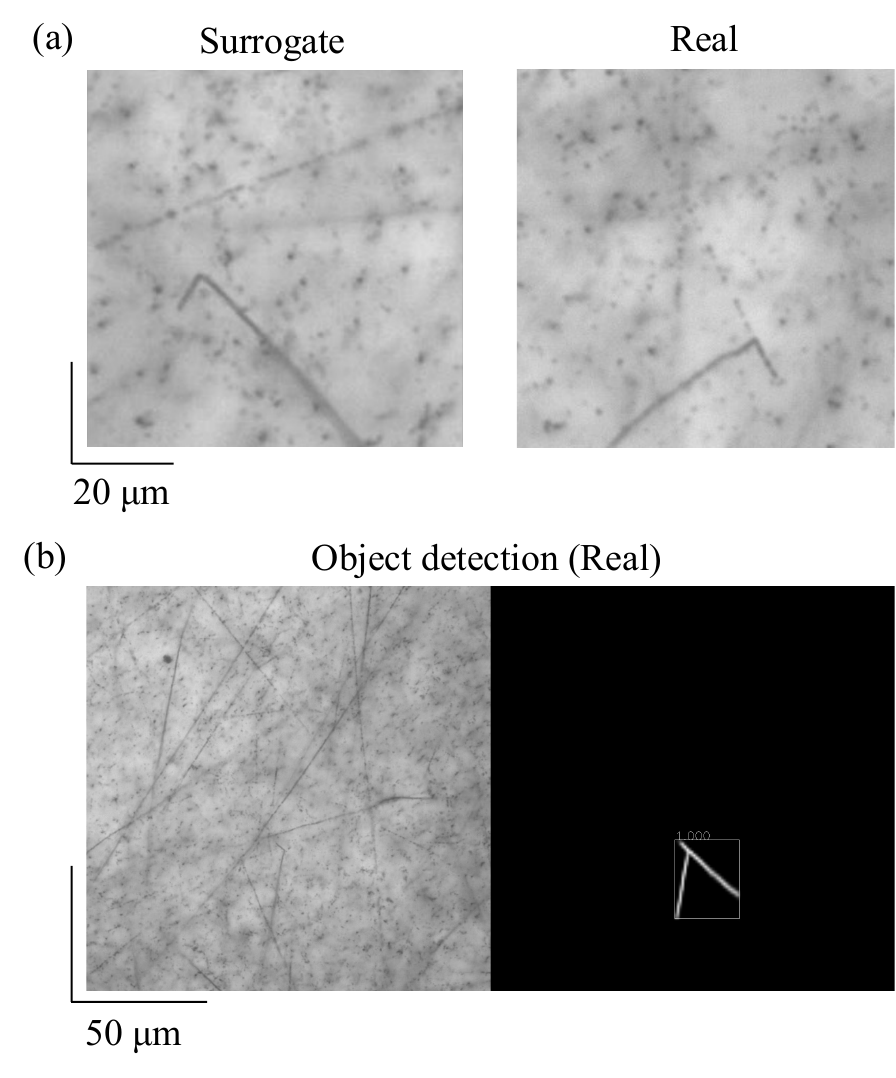}
\caption{{\bf Image generation and event detection using deep-learning.} Panel (a) shows microscopic images of a $^{3}_{\Lambda}\rm{H}$ two-body decay event generated by a deep-learning model, juxtaposed with an actual decay event recorded in the emulsion.
Panel (b) shows the result of the model’s application of object detection to a real microscopic image, accurately identifying tracks associated with a two-body decay using segmentation masks and bounding boxes.
}
\label{fig:hypertriton_detection}
\end{figure}

Figure \ref{fig:hypertriton_detection}(a) presents a surrogate image of a two-body hypernuclear decay event generated via our simulation and image-transformation, alongside a corresponding two-body decay event detected from actual emulsion data using the developed methodology.
Figure \ref{fig:hypertriton_detection}(b) shows a typical output of the object-detection model’s inference, with the left panel displaying the raw microscope image and the right panel illustrating the Mask R-CNN segmentation result overlaid on the detected track.

When Monte-Carlo samples are used to train the deep-learning model, one must ensure that the assumed values do not introduce bias into the physics analysis.
For the event search, we employed a low-magnification objective lens, which provides a wide field of view and therefore a high scanning speed. 
Consequently, the ranges of the helium track candidates of events are widely distributed, with a standard deviation of 2.7~$\rm{\mu m}$, corresponding to a $\pm 12$ MeV spread in the inferred binding energy.
This spread is more than an order of magnitude larger than the uncertainty of the binding-energy values assumed in the simulations. In addition, the binding-energy determination was performed with a separate high-magnification optical setup, and the $\pi^{-}$ track lengths were measured independently.
These two measures together strongly suppress any potential bias in the final results.

\section{Analysis and Results}
\subsection{Identification}
Under the microscope, the vertex of each candidate, corresponding to the parent hypernucleus track and its two decay tracks, was confirmed, and the track-length of the $\pi^{-}$ particle was measured.
While these tracks are traced continuously across multiple emulsion sheets, they pass through insensitive regions, such as the 40~$\rm{\mu m}$-thick base film supporting the emulsion and a 5~$\rm{\mu m}$-thick protective surface layer. 
Although the track length in these regions was rescaled to the emulsion-equivalent range based on material properties, its effect on the total path (exceeding 10 mm) was sufficiently small to be considered negligible.

\begin{figure}[ht]
\centering
\includegraphics[width=75mm]{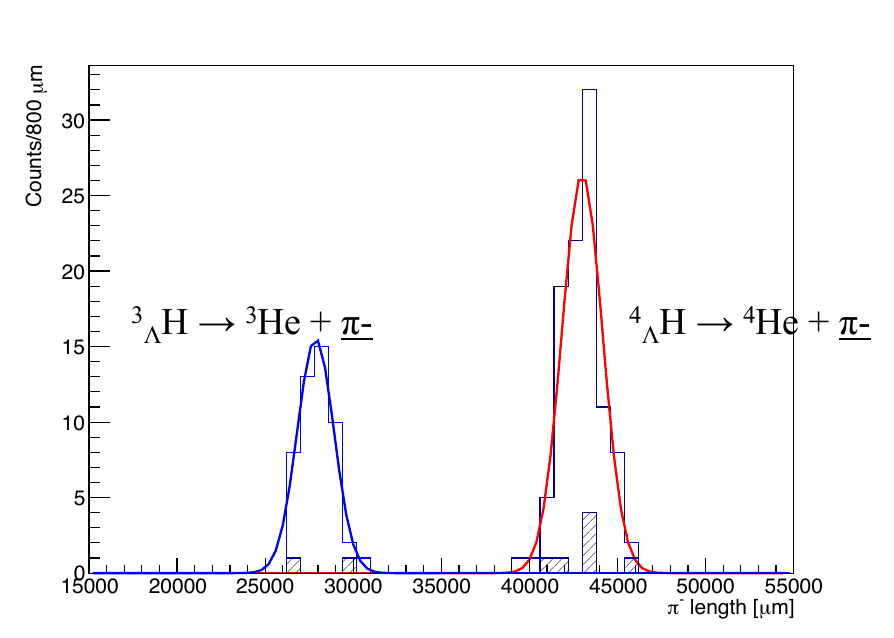}
\caption{{\bf Track-length distributions of $\pi^{-}$ from mesonic two-body decay events.}
The left and right histograms represent the track length distributions of $\pi^{-}$ particles associated with $^{3}_{\Lambda}\rm{H}$ and $^{4}_{\Lambda}\rm{H}$ decay events, respectively.
The width of these distributions corresponds to the range-straggling of $\pi^{-}$ in the emulsion.
Shaded histograms indicate the $\pi^{-}$ range excluded during the final selection of at-rest two-body decay events}
\label{fig:pi_range}
\end{figure}

Using the developed model along with manual inspection, we collected 1,213 two-body decay candidates from 0.6\% of the emulsion’s total volume.
$^{3}_{\Lambda}\rm{H}$ and $^{4}_{\Lambda}\rm{H}$ can be distinguished by the characteristic monochromatic momenta of the emitted $\pi^{-}$ particles, approximately $\sim114$ and $\sim132$MeV/$c$, respectively. 
The decay tracks of all candidates were traced within the sensitive volume of the nuclear emulsion, yielding 150 $\pi^{-}$ stationary events.
Figure \ref{fig:pi_range} shows the measured track-length distributions of $\pi^{-}$ particles from the analysis. 
These distributions clearly distinguish the decay products of $^{3}_{\Lambda}\rm{H}$ and $^{4}_{\Lambda}\rm{H}$, enabling unambiguous event identification without background event contamination.

To enhance the precision of the nuclear emulsion analysis, we performed higher-resolution imaging using optics with a pixel size of 75~nm. 
Compared with the original setup, this system provided ~3.6 times higher resolution in the $xy$-plane and 13 times higher resolution along the optical ($z$) axis. 
To minimize human bias in measuring track-lengths and decay points, we employed an automated procedure that combined image processing, brightness fitting, and a Kalman filter based on the Genfit library \cite{Ekawa_D, GenFit}.

Two criteria were used to identify ‘at-rest’ two-body decays. 
First, the trajectories of the parent nucleus and its two daughter particles were required to lie in a single plane within a 3$\sigma$ tolerance. 
Second, the inner product of the angles between the two daughter particles had to be negative (also within a three $\sigma$). 
Applying these criteria, together with the accurately measured range of helium isotopes from the decay, enabled us to identify 46 $^{3}_{\Lambda}\rm{H}$ and 95 $^{4}_{\Lambda}\rm{H}$ events consistent with two-body decays at rest. 
These events were subsequently used for binding energy calculations.

\subsection{Range-Energy calibration}
We then calculated their invariant masses via momentum analysis of daughter particles.
In conventional kinetic analyses using nuclear emulsions, a particle’s kinetic energy is determined by converting its range, the total track length to the stopping point, into energy using the range–energy relation
\begin{align*}
    R &= \frac{M}{Z^{2}}\lambda(\beta) + R_{ext} \\
    R_{ext} &= MZ^{2/3} C_{Z} (\beta / Z)\\
    \frac{\lambda_{s}}{\lambda} &= \frac{rd - 1}{rd_{s} - 1} + \frac{r(d_{s} - d)}{rd_{s} - 1} \frac{\lambda_{s}}{\lambda_{w}} 
\end{align*}
where $R$, $Z$, and $M$ denote particle range, charge, and  mass in proton mass units, respectively \cite{Barkas, Heckman, Barkas1958}.
$R_{ext}$ signifies effective charge reduction due to electron capture, while $C_{Z}$ is an empirical correlation coefficient dependent on $\beta/Z$.
$\lambda$ represents the track length of a proton with velocity $\beta$, and $\lambda_s$ and $\lambda_w$ correspond to the ranges in standard emulsion and water, respectively.
Here, $d$ and $d_s$ are the density parameter of the emulsion under analysis and the density of the standard emulsion, respectively.
The factor $r$ corrects for the increase in the volume-to-mass ratio owing to water absorption.
Originally formulated by Barkas et al. in the 1960s using Ilford's standard emulsion, this equation has been widely used in subsequent studies \cite{Barkas}.

The density of a nuclear emulsion is highly sensitive to its composition and moisture content, rendering it a crucial parameter for adapting the range–energy relationship to emulsion types that differ from the standard emulsion.
However, this density parameter does not directly represent the physical material density, but serves as a scaling factor.
For the nuclear emulsions used in the KEK/J-PARC experiments, $d$ was determined by measuring the range of monoenergetic $\alpha$ particles emitted by naturally occurring radioactive elements within the emulsion gelatin.

We collected $\alpha$-particles from a $10 \times 10 \times 0.05~\rm{cm}^{3}$ region surrounding each hypernuclear event to calibrate the density parameters for calculating the momenta of helium isotopes and $\pi^{-}$ mesons from two-body decays.
However, when applying this conventional method, inconsistencies arose.
In the decay $^{4}_{\Lambda}\rm{H} \rightarrow {}^{4}\rm{He} + \pi^{-}$, the $^{4}\rm{He}$ momentum was measured as $132.96 \pm 0.38$~MeV/$c$, whereas the $\pi^{-}$ momentum was $134.52 \pm 0.14$~MeV/$c$.
Despite selecting back-to-back emissions from at-rest two-body decay, a $3.9\sigma$ deviation was observed.
Given that the range-energy relation was calibrated using $\alpha$ particles and that the measured $^{4}\rm{He}$ momentum agreed with previous reports of $^{4}_{\Lambda}\rm{H}$ decay, we conclude that the $\pi^{-}$ momentum is systematically overestimated \cite{MAMI_2016}.

\begin{figure}[ht]
\centering
\includegraphics[width=70mm]{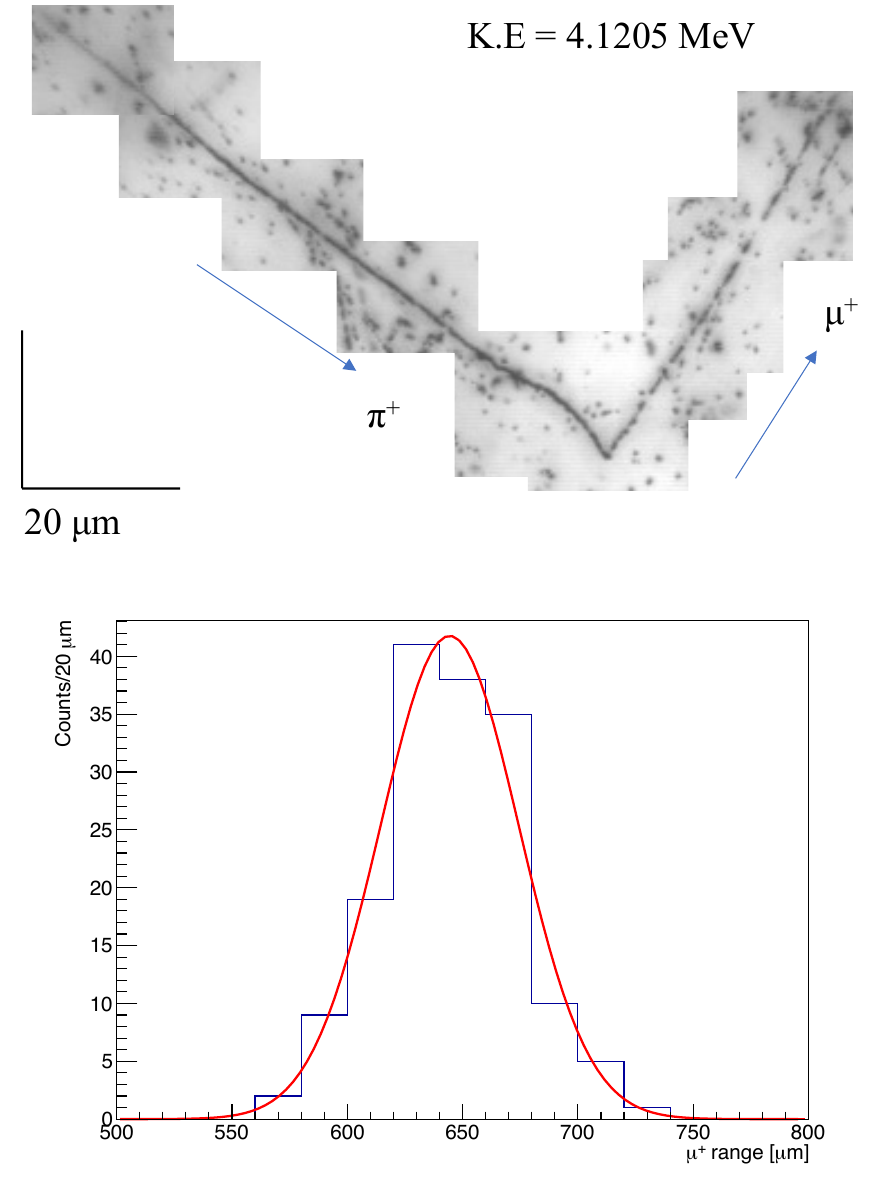}
\caption{{\bf Microscopic images of $\pi^{+}$ at rest and its decay, along with the range distribution of monoenergetic $\mu^{+}$.}
The $\pi^{+}$ produced by the irradiated $K^{-}$ beam gradually lost energy as it passed through the emulsion detector.
After coming to rest, $\mu^{+}$ was released with a kinetic energy of 4.1205~MeV following the decay process of $\pi^{+} \rightarrow \mu^{+} + \nu_{\mu}$.
The track-length of the corresponding particle in the E07 nuclear emulsion sheet was calculated to be $644.4 \pm 2.4$~$\rm{\mu m}$, and the density parameter was determined to be $3.379 \pm 0.016~\rm{g/cm^{3}}$ based on this energy and the ATIMA stopping power calculation results.
}
\label{fig:pi_mu}
\end{figure}

To further investigate this issue, we introduced an alternative calibration sample utilizing the range of $\mu^{+}$ particles produced via $\pi^{+} \rightarrow \mu^{+} + \nu_{\mu}$ weak decay. 
Figure. \ref{fig:pi_mu} shows microscopic image of $\pi^{+}$ decay event and range distribution of $\mu^{+}$ tracks from such decay events in nuclear emulsion. 
These particles had a monochromatic energy of 4.1205 MeV (29.794 MeV/$c$) and traveled ~640~$\rm{\mu m}$ in the emulsion.
Traditionally, identifying $\pi^{+}$ decay events requires manual visual inspection or basic image processing, limiting the number of detectable events.
By leveraging an object detection model originally developed for hypernuclear searches, we achieved a detection efficiency four times higher than that of visual scanning and twice that of conventional image processing.
Using the conventional range–energy relation, we derived density parameters of $3.516 \pm 0.016~\rm{g/cm^{3}}$ and $3.616 \pm 0.011~\rm{g/cm^{3}}$ from $\mu^{+}$ and $\alpha$ particles, respectively.
These values differed by $5\sigma$, even when collected from the same emulsion volume, indicating that the conventional formula failed to treat particles with different charges and masses in a consistent manner.

These findings indicate that the conventional range–energy relationship does not consistently apply to all particles in the E07 nuclear emulsion.
For Ilford's standard emulsion, empirical functions such as $C_{z}$ enabled the range–energy relation to be effectively adapted across different particle species, with density serving as a key calibration parameter.
However, the E07 nuclear emulsion differs significantly in composition, preventing the uniform treatment of particles with widely varying charges and masses, even when applying the density parameter.
Furthermore, functions such as $C_{z}$, originally derived from the experimental results of Barkas et al., do not account for variations in emulsion composition as input parameters.
This limitation renders modifying the conventional range–energy relation impractical for direct application in our analysis.

To address these limitations, we employed ATIMA v1.41, a modern stopping-power calculation tool developed in 2018 \cite{ATIMA, helmut}.
Unlike empirical formulations such as $C_{z}$, ATIMA incorporates the elemental composition of materials, enabling more precise calculation of energy loss and range for different particles.
However, ATIMA does not support stopping-power calculations for $\mu^{+}$ and $\pi^{-}$, necessitating an independent evaluation of its accuracy.
First, we used hydrogen ions as a reference and evaluated ATIMA’s performance by comparing its output with stopping-power data from previous studies.
The pions under investigation possess a kinetic energy of ~50 MeV; therefore, protons at ~30 MeV served as a suitable reference, given their similar $\beta \gamma$.
In this energy region, the discrepancy in $dE/dx$ between experimental data and the predecessor of SRIM–ATIMA was ~0.1\% \cite{SRIM_accuracy}.
Although ATIMA may improve the accuracy by incorporating more recent data, we adopted the ~0.1\% accuracy as the baseline for systematic uncertainty in range–energy conversion.
Furthermore, the discrepancy between measured and modeled $dE/dx$ increases in the lower-energy regime; this effect is considered in our uncertainty estimates during density calibration with low energy $\mu^{+}$ particles.

Secondly, to further validate the ATIMA-based calculations for $\pi^{-}$, we investigated additional effects that could influence stopping power, namely, bremsstrahlung and the Barkas effect \cite{Barkas_effect, Barkas_effect_proton}.
Pions are significantly lighter than protons; therefore, bremsstrahlung was initially considered, as it contributes ~0.3\% of the electron-stopping power in the studied energy range \cite{estar}.
The bremsstrahlung cross section $\sigma$ scales with the inverse square of the particle mass, $\sigma \propto \left(\frac{e^{2}}{mc^{2}}\right)^{2}$.
Given the mass ratio of $\pi^{-}$ to an electron (approximately 273:1), the expected bremsstrahlung effect for $\pi^{-}$ is negligible, in the order of $10^{-6}$.
Next, we examined the reduced stopping power of negatively charged particles relative to their positively charged counterparts of the same mass, a phenomenon known as the Barkas effect.
For a range $R \approx 200$~$\rm{\mu m}$, negatively charged pions travel $\Delta R \approx 6$~$\rm{\mu m}$ farther than positively charged pions of the same energy \cite{Barkas_effect_pi}.
\begin{align*}
\Delta R = (6 \pm 0.3)[1-\exp(-\frac{R-1.1}{45 \pm 10})]
\end{align*}
The resulting difference in kinetic energy was $-43 \pm 2$~keV.
We adopted this average value as a correction to the ATIMA calculations.

By applying ATIMA in conjunction with low-energy $\mu^{+}$ particles, we determined the density parameter to be $3.379 \pm 0.016~\rm{g/cm^{3}}$, which agrees well with the actual density of $3.40 \pm 0.04~\rm{g/cm^{3}}$ obtained from volume and weight measurements.
Additionally, we evaluated the sheet-to-sheet variation in the density parameter using $\alpha$ particles, yielding an uncertainty of $\pm~0.006$, which was included as a systematic error.
Applying this revised density parameter, we calculated the momentum of $\pi^{-}$ particles emitted from the at-rest two-body decay of $^{4}_{\Lambda}\rm{H}$ as $p_{\pi^{-}} = 132.761 \pm 0.131(\mathrm{Stat.}) \pm 0.086(\mathrm{Syst.})$~MeV/$c$.
This result is consistent with the back-to-back helium momentum emitted from $^{4}_{\Lambda}\rm{H}$ decay events.
Thus, by leveraging ATIMA-based stopping-power calculations and $\mu^{+}$-based calibration, we successfully established a reliable method for determining $\pi^{-}$ momentum in the E07 nuclear emulsion.

\subsection{Binding energy}
\begin{figure}[ht]
\centering
\includegraphics[width=70mm]{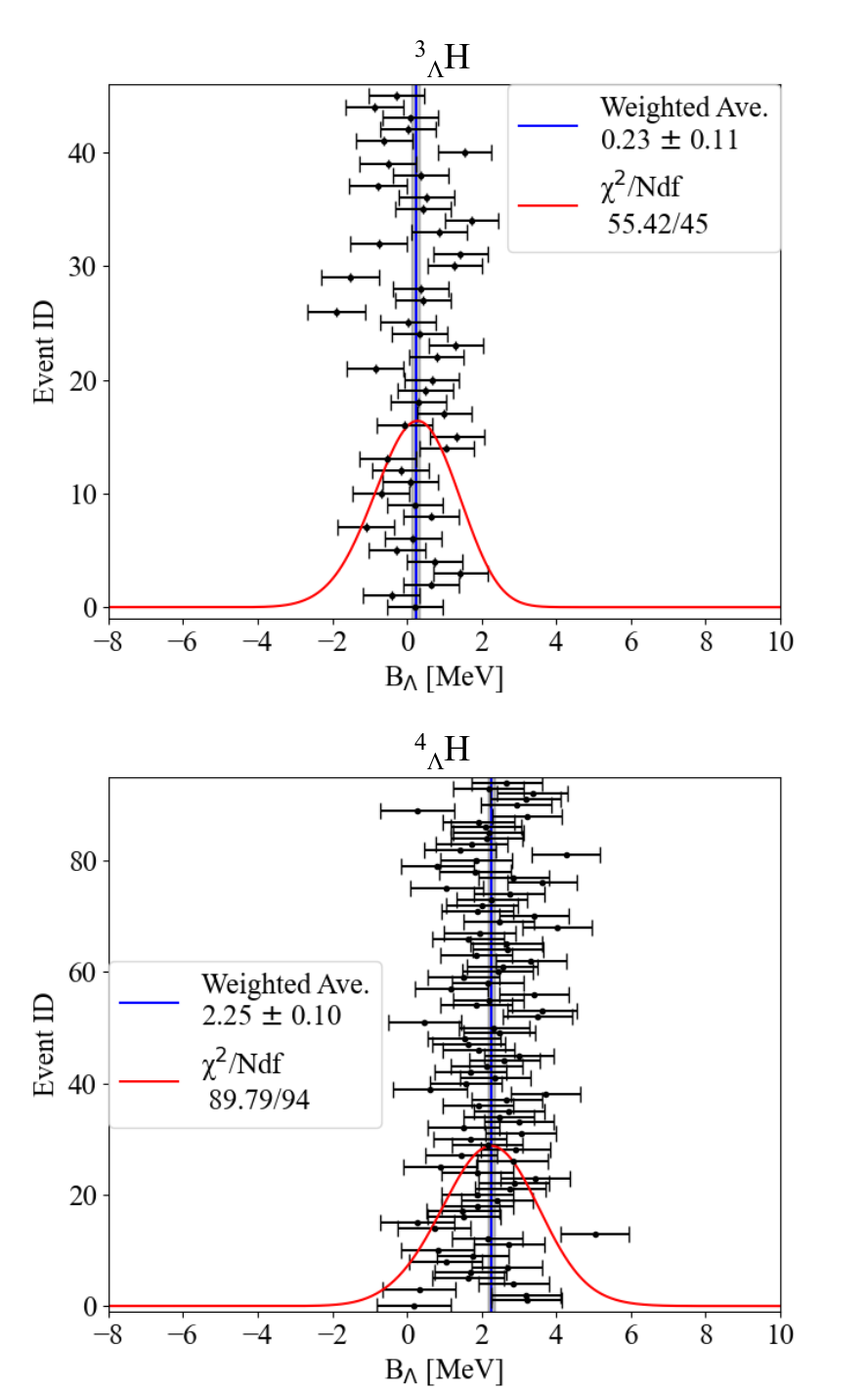}
\caption{{\bf Results of $\Lambda$ binding energies for $^{3}_{\Lambda}\rm{H}$ and $^{4}_{\Lambda}\rm{H}$.} Each point corresponds to the binding energy calculated from an individual event, with error bars reflecting uncertainties in track lengths, including range straggling.
The vertical axis indicates the event ID.
The red curve depicts the overall probability distribution, whereas the blue band represents the statistical error range.}
\label{fig:binding_energy}
\end{figure}

From the calculated momentum of $\pi^{-}$, the binding energies of the $\Lambda$ particles in the hypernucleus were deduced using the following equation:
\begin{align*}
M(^{3}_{\Lambda}\rm{H},~^{4}_{\Lambda}\rm{H}) =& \sqrt{M^{2}(\rm{^{3,~4}He}) + \lparen \it p_{\pi^{-}}/c\rparen^{\rm{2}}} + \sqrt{M_{\pi-}^{2} + \lparen \it p_{\pi^{-}}/c\rparen^{\rm{2}}}\\
B_{\Lambda} =& \lbrace M(\rm{^{2,~3}H}) + M_{\Lambda} - M(^{3}_{\Lambda}\rm{H},~^{4}_{\Lambda}\rm{H})\rbrace \it c^{\rm 2}.
\end{align*}
In our calculation, we selected only at-rest two-body decays. 
We omitted momentum measurements of helium isotopes with short track lengths, which were affected by straggling and detector sensitivity limitations.
As nuclear emulsion–based identification and subsequent analysis can be treated as independent measurements, we employed a weighted average of the results to determine the binding energies (See Figure \ref{fig:binding_energy}).
The results are
\begin{align*}
(^{3}_{\Lambda}\rm{H})~B_{\Lambda} = 0.23 \pm 0.11 \rm{(Stat.)} \pm 0.05 \rm{(Syst.)}~\rm{MeV}\\
(^{4}_{\Lambda}\rm{H})~B_{\Lambda} = 2.25 \pm 0.10 \rm{(Stat.)} \pm 0.06 \rm{(Syst.)}~\rm{MeV}
\end{align*}
The effects of various factors on systematic errors in the binding energy measurements are presented in Table~\ref{tab:syst_error}.

\begin{table}[ht]
    \centering
    \caption{{\bf Summary of systematic uncertainties in the binding energy measurements}\label{tab:syst_error}}
    \begin{tabular}{c|c|c|c}
    &Ref.&$^{3}_{\Lambda}\rm{H}$ [keV]&$^{4}_{\Lambda}\rm{H}$ [keV]\\
    \hline
    Mass of $\Lambda$&[\cite{PDG}]&6&6\\
    \hline
    Mass of $\rm{He}$ &[\cite{PDG}]&$8.5 \times 10^{-3}$&$1.1 \times 10^{-3}$\\
    \hline
    Mass of $\rm{^{2}H}$&[\cite{PDG}]&$5.7 \times 10^{-4}$&-\\
    \hline
    Mass of $\rm{^{3}H}$ &[\cite{PDG}] &-&$8.5 \times 10^{-4}$\\
    \hline
    Mass of $\pi^{-}$ &[\cite{PDG}]&$8.5 \times 10^{-3}$&$8.5 \times 10^{-3}$\\
    \hline
    $dE/dx$ Exp./Theory $\pi$&[\cite{SRIM_accuracy}]&40&50\\
    \hline
    Barkas effect for $\pi^{-}$&[\cite{Barkas_effect_pi}]&2&2\\
    \hline
    Density calibration&-&33&33\\
    \hline
    Total&&$\sim$50&$\sim$60
    \end{tabular}
\end{table}

\section{Conclusion}
The application of deep-learning to the latest nuclear emulsion data enabled the detection of single-$\Lambda$ hypernuclear events, although this was not the original objective of the experiment.
Simultaneously, it shed light on and quantitatively addressed a longstanding issue with the range–energy equation.
Historically, nuclear emulsion experiments in the 1960s and 1970s reported an accuracy of ~40 keV for binding energy measurements \cite{Davis}, with the present study attaining a similar level of precision.
Although this remains slightly above the 30~keV simulation limit, which assumes a perfectly calibrated range-energy equation \cite{Liu2021}—further improvements can be realized by establishing dedicated range–energy relations for each particle type.
By applying particle-specific energy calibrations, we can carry out precision analyses of three-body decay events in both single-$\Lambda$ and double-$\Lambda$ hypernuclei.
In these decays, the $\pi^{-}$ track lengths, which dominate the mass resolution, are shorter than in two-body channels, so the systematic uncertainties arising from the stopping-power ambiguities are then further suppressed.
Furthermore, the statistical uncertainties on the binding energies will be reduced through additional analysis of two-body decay events and incorporating the three-body decay channel, since the current results are based on only 0.6\% of the total E07 data set.

\begin{figure}[ht]
\centering
\includegraphics[width=70mm]{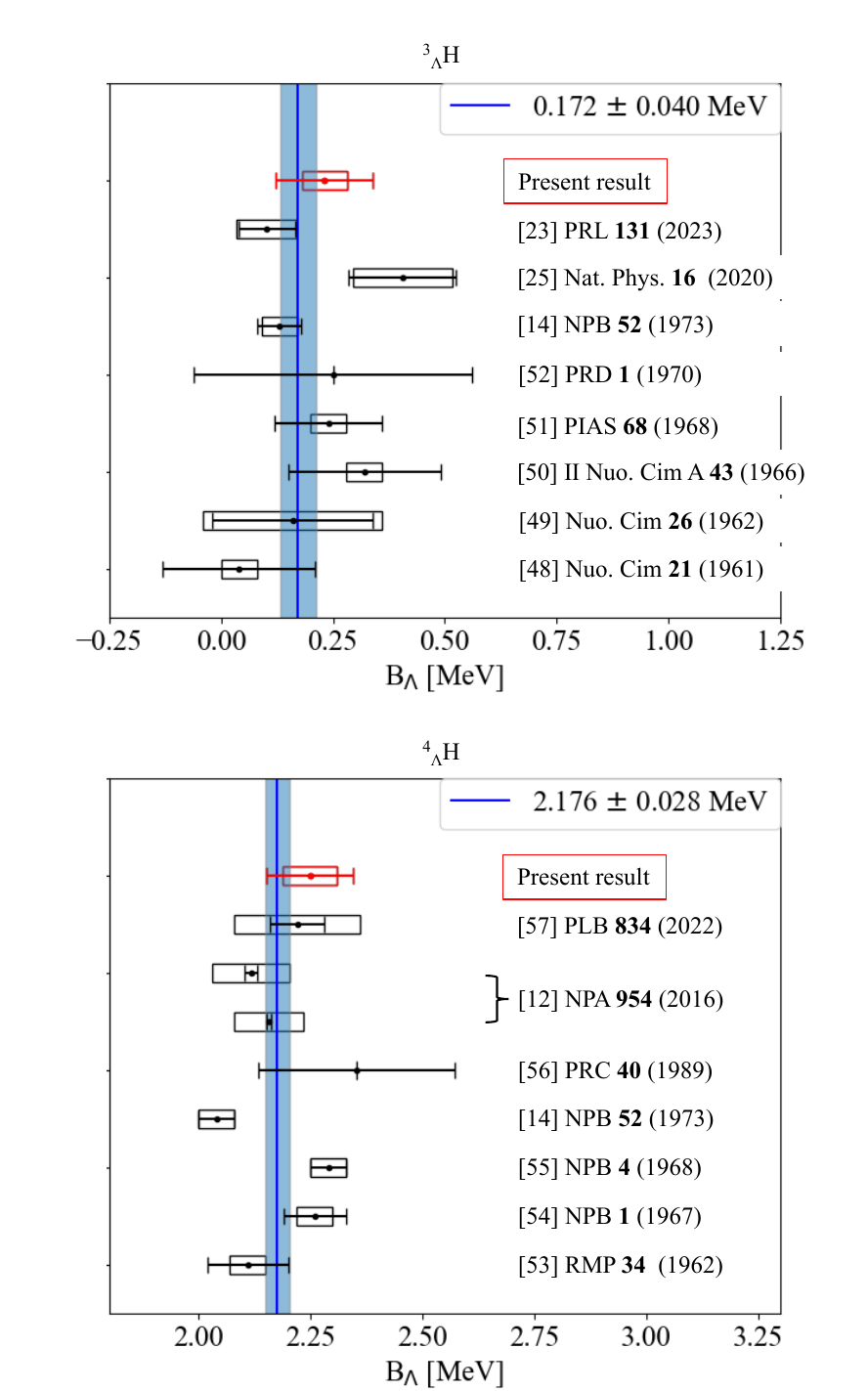}
\caption{{\bf World average of binding energies for $^{3}_{\Lambda}\rm{H}$ and $^{4}_{\Lambda}\rm{H}$.}
This includes the results from the present study alongside previous measurements \cite{Prakash1961, Ammae1962,Mayeur1966ADO, Chaudhari1968, PhysRevD.1.66, STAR_2020, ALICE_2023, RevModPhys.34.186, GAJEWSKI1967105, BOHM1968511, Juric, PhysRevC.40.R479, MAMI_2016, STAR_A4}.
The bars and boxes in the plots indicate statistical and systematic uncertainties, respectively.}
\label{fig:world_average}
\end{figure}

The weighted average of existing world data yields binding energies of $0.172 \pm 0.040$~MeV for $^{3}_{\Lambda}\rm{H}$ and $2.176 \pm 0.028$~MeV for $^{4}_{\Lambda}\rm{H}$, consistent with our findings (see Figure \ref{fig:world_average}).
The loosely-bound nature of hypertritons has also been directly observed in data from heavy-ion collisions and emulsion-based studies, including the present work \cite{Juric, STAR_2020, ALICE_2023}.
In recent years, growing interest in the “hypertriton puzzle” has spurred the rapid development of indirect methods for determining binding energies.
For instance, the STAR collaboration extracted $B_{\Lambda}$ values via particle correlation and femtoscopy analyses involving $\Lambda$ particles and deuterons \cite{dlambda_star, dlambda_star_2}.
It estimated $B_{\Lambda} = 0.04\mathrm{-}0.33$~MeV at a 95\% confidence level.
By improving the precision of direct invariant-mass measurements, femtoscopic correlation analyses of $\Lambda-p$ and $\Lambda-d$, and theoretical calculations, a more comprehensive understanding of three-body $\Lambda pn$ forces and nuclear medium effects in a hypertriton system can be achieved.

The $^{4}_{\Lambda}\rm{H}$ result merits further investigation.
Juric et al. identified a systematic error in the range–energy relation for $\pi^{-}$ tracks exceeding 4~cm, based on invariant mass analyses using a $\Lambda$-hyperon calibration source. Consequently, only three-body decay events were eventually included \cite{Bohm1970, Juric}.
Although their revised values were among the most precise of the era, they deviated from the global average.
In contrast, Gajewski et al. and Bohm et al. reported $^{4}_{\Lambda}\rm{H}$ binding energies of $2.26 \pm 0.07 \rm{(Stat.)} \pm 0.04 \rm{(Syst.)}$~MeV and $2.29 \pm 0.04 \rm{(Stat.)} \pm 0.04\rm{(Syst.)}$~MeV, respectively, \cite{GAJEWSKI1967105, BOHM1968511}—which closely match our findings.
Similarly, decay-pion spectroscopy from MAMI-C (based on two-body decay measurements) indicates a slightly larger binding energy than that reported by Juric et al. \cite{MAMI_2016}.
Our analysis does not rely on the conventional range–energy formula; instead, low-energy $\mu^{+}$ calibration is used to suppress stopping-power errors in the relevant energy region. 
This suggests that the systematic errors noted in 1970 for high-momentum $\pi^{-}$ may have stemmed from uncertainties associated with other particles, particularly in three-body decay processes.
Notably, at that time, the $^{3}_{\Lambda}\rm{H}$ result was obtained from a combination of analyses involving two-body and three-body decay data.

When the data from three-body decay analyses reported in Ref.\cite{Juric, RevModPhys.34.186} are excluded, the weighted average of the $^{4}_{\Lambda}\rm{H}$ binding energy becomes $2.23 \pm 0.03$~MeV.
The effect of charge-symmetry breaking (CSB) in $\Lambda N$ interactions was initially quantified at $\Delta B_{\Lambda} = 350$~keV, reflecting the binding energy difference between $^{4}_{\Lambda}\rm{H}$ and $^{4}_{\Lambda}\rm{He}$ measured in early nuclear emulsion experiments \cite{Juric}.
However, using $^{4}_{\Lambda}\rm{H}$ binding energy derived exclusively from two-body decays yields $\Delta B_{\Lambda} \approx 160$~keV, suggesting that CSB in the ground state may be smaller for $A=4$ hypernuclei.
Looking ahead, new measurements of $^{4}_{\Lambda}\rm{H}$, along with upcoming $^{4}_{\Lambda}\rm{He}$ data acquired using the method established here, as well as gamma-ray spectroscopy from excited states and decay-pion spectroscopy, will collectively advance our understanding of CSB in the $\Lambda N$ interaction, a key feature of baryonic interactions involving strangeness \cite{CSB_Dalitz, CSB_Gal, MAMI_2016, Yamamoto, STAR_2022}.

Based on this newly established energy calibration, previously reported binding energies of double-strangeness hypernuclear events may require revision.
Notably, the double-$\Lambda$ hypernuclear event in $^{~~6}_{\Lambda\Lambda}\rm{He}$ (known as the NAGARA event) remains the only uniquely identified event involving mesonic decay, and continues to serve as an essential benchmark for studying interactions in the $S=-2$ sector \cite{NAGARA}.
The results of this re-analysis will be reported in another publication.

\clearpage
\section*{Data and Code Availability}
Raw data were generated at the High Energy Nuclear Physics Laboratory, RIKEN. They are available from the corresponding author upon reasonable request.
The measurement data and the computer codes to generate results are available at \url{https://github.com/ayumisalt/Rangeenergy_E07}.

\section*{Acknowledgment}
This work was supported by JSPS KAKENHI through Grants JP16H02180, JP20H00155, JP18H05403, JP19H05147 and JP25H00404 (Grants-in-Aid for Scientific Research on Innovative Areas 6005). 
The authors acknowledge support from Proyectos I+D+i 2020 (ref: PID2020-118009GA-I00); Proyectos I+D+i 2022 (ref: PID2022-140162NB-I00); and Proyecto Consolidación Investigadora 2022 (ref:CNS2022-135768); as well as from Grants 2019-T1/TIC-13194 and 2023-5A/TIC-28925 under the Atracción de Talento Investigador program of the Community of Madrid.
H.W. acknowledges support from the Major Science and Technology Projects in Gansu Province under Grant No. 24GD13GA005. 
A. K. was supported by JSPS KAKENHI Grant No. JP23K19051, JP25H01550 and JP25K17415. (Grant-in-Aid for Research Activity Start-Up, Grant-in-Aid for Transformative Research Areas and Grant-in-Aid for Early-Career Scientists). 
The authors thank the J-PARC E07 collaboration for providing the emulsion sheets.
The authors thank Dr. Hiroki Rokujo and the J-Science Group for their assistance in analyzing the compositions of the emulsions, and 
Dr. Helmut Weick for his detailed advice on ATIMA.
We also thank Michi Ando, Chiho Harisaki, Risa Kobayashi, and Hanako Kubota of RIKEN, and Yoko Tsuchii of Gifu University, for their technical support during the event mining in the J-PARC E07 nuclear emulsions. 
The authors also thank Yukiko Kurakata of RIKEN for her administrative work.

\vspace{0.2cm}
\noindent

\let\doi\relax

\clearpage
\bibliographystyle{ptephy}
\bibliography{2505-035-3D-AyumiKasagi}

\end{document}